\documentclass{jnmp}

\begin{document}

\renewcommand{\evenhead}{G~Gaeta,~S~Walcher}
\renewcommand{\oddhead}{Dimension increase and splitting}

\def\giorno{4/9/2004}

\def\pa{\partial}
\def\lapl{\triangle}
\def\grad{\nabla}
\def\ss{\subset}
\def\xd{{\dot x}}
\def\yd{{\dot y}}
\def\xb{{\bf x}}
\def\vb{{\bf v}}
\def\eb{{\bf e}}
\def\pb{{\bf p}}

\def\D{\Delta}
\def\a{\alpha}
\def\b{\beta}
\def\ga{\gamma}
\def\eps{\varepsilon}
\def\la{\lambda}
\def\th{\vartheta}
\def\d{{\rm d}}
\def\de{\delta}
\def\De{\Delta}
\def\s{\sigma}
\def\vth{\vartheta}
\def\om{\omega}
\def\z{\zeta}
\def\F{{\cal F}}
\def\R{{\bf R}}
\def\({\left(}
\def\){\right)}
\def\[{\left[}
\def\]{\right]}
\def\<{\langle}
\def\>{\rangle}
\def\.#1{{\dot #1}}
\def\^#1{{\widehat #1}}
\def\t#1{{\widetilde #1}}

\thispagestyle{empty}

\FirstPageHead{12}{S1}{2005}{327--342}{Article}

\copyrightnote{2005}{G~Gaeta,~S~Walcher}

\Name{Dimension increase and splitting for \\ Poincar\'e-Dulac normal forms}

\Author{Giuseppe GAETA}
\Address{Dipartimento di Matematica, Universit\`a di Milano, I-20133 Milano, (Italy)\\
E-mail: gaeta@mat.unimi.it}

\Author{Sebastian WALCHER}
\Address{Lehrstuhl A f\"ur Mathematik, RWTH Aachen, D-52056 Aachen
(Germany) \\ E-mail: walcher@mathA.rwth-aachen.de}

\Date{ }

\begin{abstract}
\noindent Integration of nonlinear dynamical systems is usually
seen as associated to a symmetry reduction, e.g. via momentum map.
In Lax integrable systems, as pointed out by Kazhdan, Kostant and
Sternberg in discussing the Calogero system, one proceeds in the
opposite way, enlarging the nonlinear system to a system of
greater dimension. We discuss how this approach is also fruitful
in studying non integrable systems, focusing on systems in normal
form.
\end{abstract}

\section{Introduction. Dimension increase and splitting}

Finite dimensional integrable systems have been a rarity for a
very long time; up to few decades ago, they could be easily
enumerated: harmonic oscillators, the Kepler system, a few
spinning tops.

Apart from harmonic oscillators -- which give linear equations and are the prototype of integrability -- these are integrated by exploiting their symmetries and the
associated integrals of motion to reduce the problem to a lower
dimensional one; thus, we have {\it symmetry reduction}.

The situation changed radically in the seventies, when entire
classes of new integrable systems appeared, integrable by the Lax
pair construction \cite{Lax}. Among these a
prominent role -- also due to priority -- is taken by the Calogero
system \cite{Cal}. In this case, one has an arbitrary number $N$
of points on the line, interacting via a certain  pair potential;
thus the system is described by a natural hamiltonian $H = (1/2)
\sum_i (\dot{x}_i )^2 + V (x)$. Calogero showed that there is an
invertible map from $\R^N$ to $GL(N)$, mapping the vector $x$ to a
matrix $L(x)$, such that the evolution of $x (t)$ results in an
evolution of $L (t) := L [x(t)]$ governed by Lax equations; thus
one integrates $L(t)$ and obtains  $x(t)$.

As pointed out by Kazhdan, Kostant and Sternberg \cite{KKS}, this
also initiated a new way of integrating systems (which they
studied geometrically): indeed, rather than trying to reduce the
dimension of the system, Calogero passed to consider a system of
{\it higher dimension}\footnote{Another classical
instance of simplification by dimension increase is the trick to
linearize the matrix Riccati equation. See also \cite{CGM,ShW} in
this context.}. In this note we want to show that this approach is fruitful also beyond the realm of integrable hamiltonian systems.

The key observation made by Kazhdan, Kostant and Sternberg
\cite{KKS}, see also \cite{Mar,MSSV}, in the context of
Hamiltonian systems is that given a differential equation $\Delta$
in dimension $n$, in some cases it may be helpful to increase its
dimension: this is in particular the case when it is possible to
describe $\Delta$ as originating from either the symmetry
reduction or the projection to an $n$-dimensional manifold of a
simpler (e.g. linear) equations $\Delta^*$ in dimension $n+m$.

It was remarked in \cite{con} that this approach does apply also
to non-hamiltonian systems, and in particular to systems in
Poincar\'e-Dulac normal form \cite{Arn,CiGa,enf} (a remark to this effect was already contained in \cite{enf}). Essentially, a system
in normal form with only a finite number of resonances can
always be mapped to a linear system, and is equivalent to the
latter on a certain invariant manifold. As observed there, such a
procedure is geometrically interesting, but equivalent
to a classical technique already known to Dulac \cite{Dul}, and
abscribed by him to Horn and Lyapounov. In more general terms,
i.e. outside the scope of normal forms theory, this corresponds to
the situation of a finite dimensional centralizer for a certain
algebra associated to the vector field \cite{cag}.

In the present note, we show how ``dimension increasing''
nicely combines with a ``splitting'' (in a sense to be described
below) approach \cite{split} in the case of an infinite
number of resonances\footnote{For a different use of the splitting approach in normal forms type problems, see \cite{YP}.}; this case cannot be tackled by the Horn--Lyapounov--Dulac approach.

Roughly speaking, the procedure presented in this note combines
the two approaches mentioned above, i.e. describing a nonlinear
system as projection of a linear one and symmetry reduction in the
sense described in \cite{split}.

A more abstract (and general) treatment is given elsewhere \cite{seb}; in the present note we adopt an approach and notation aimed at
applications, and discuss a number of concrete applications. We
focus on systems in normal form (sect.2) and in particular on the
embedding and splitting of systems with infinitely many resonances (sect.3). Some simple examples are discussed in sect.4, while sect.5 is devoted to (in general non-hamiltonian) nonlinearly perturbed oscillators, and sect.6 to bifurcation problems.

\section{Equations in normal form; resonances}

We consider an ODE in $R^n$ with a fixed point in the origin and
expanded around this in a power series; we write this in the form
$$ {\d x \over \d t } \ = \ f (x) \ = \ A x \, + \, \sum_{k=1}^\infty f_k (x) \eqno(1) $$
where $f_k$ is polynomial with $f_k (a x) = a^{k+1} (x)$, and we have singled out the
linear part $f_0 (x) = A x$. We also write $F$ for the nonlinear
part of $f$, i.e. $f^i (x) = A^i_j x^j + F^i (x)$.

As well known, the matrix $A$ can be uniquely decomposed into its
semisimple and nilpotent parts, which commute with each other and
hence with $A$:
$$ A = A_s + A_n \ ; \ [A_s , A_n ] = 0 . $$
In the following, we will use the following vector fields ($\pa_i
:= \pa / \pa x^i$):
$$ X_A := (Ax)^i \pa_i \ , \ X_0 = (A_s)^i \pa_i \ , \ X_F = F^i (x) \pa_i \ , \ X_f = f^i (x) \pa_i \ . $$

\subsection{Resonances}

Let us denote by $\{ \la_1 , ... , \la_n \}$ the eigenvalues of
$A$; take a basis $\{ \eb_1 , ... , \eb_n \}$ in $R^n$ consisting
of generalized eigenvectors of $A$, i.e. eigenvectors of $A_s$: $A_s \eb_j = \la_j \eb_j$. We will use $x$ coordinates in this basis, and the multiindex notation
$$ x^\mu \ := \ x_1^{\mu_1} ... x_n^{\mu_n} \ . $$

We say that the vector monomial $\vb_{\mu,\a} := x^\mu \eb_\a$ is {\it
resonant} with $A$ if
$$ (\mu \cdot \la) \ := \sum_{i=1}^n \mu_i \la_i \ = \ \la_\a \ \
{\rm with} \ \ \mu_i \ge 0 \ , \ |\mu| := \sum_{i=1}^n \mu_i \ge 1 \ . \eqno(2) $$

The relation $(\mu \cdot \la) = \la_\a$ is said to be a {\it
resonance relation} related to the eigenvalue $\la_\a$, and the
integer $|\mu|$ is said to be the {\it order} of the resonance.
In our context it is useful to include order one resonances in the definition (albeit the {\em trivial} order one resonances given by $\lambda_\alpha = \lambda_\alpha$ are of little interest).
Note that here one could as well consider $A_s$ rather than $A$.

The space of vectors resonant with (the semisimple part of) $A$ is
defined as the linear span of the vectors $\vb_{\mu,\a}$ defined
above.

\subsection{Normal forms}

We say (see e.g. \cite{Arn,CiGa,enf}) that (1) is in
Poincar\'e-Dulac normal form\footnote{The reader should be warned that some different definition is also in use \cite{Elp}.} if its nonlinear part $F(x)$ is resonant with $A$. This implies that
$$ \[ \, X_0 \, , \, X_F \, \] \ = \ 0 \ . \eqno(3) $$

It should be mentioned that the presence of a nilpotent part $A_n$
in $A$ introduces some subtleties. If (3) holds, then both $X_A$ and $X_F$ commute with $X_0$, and therefore
$$ \[ \, X_0 \, , \, X_f \, \] \ = \ 0 \ , \eqno(4) $$
i.e. the system has a symmetry described by a semisimple matrix.

As well known, starting from any dynamical system (or vector
field)  of the form (1), we can arrive at a dynamical system (or
vector field) in Poincar\'e-Dulac normal form by means of a
sequence (in general, infinite) of near-identity transformations 
obtained by means of the Poincar\'e algorithm; these combine into
a near-identity transformation $H$ defined by a series which is in
general only formal. 

\medskip\noindent
{\bf Remark 1.} We may reformulate the definition of systems in
normal form by saying that $f$ is is normal form if and only if
$X_F$ is in the centralizer of $X_0$. $\odot$

\medskip\noindent
{\bf Remark 2.} If the system is required to have some symmetry,
say $[X_f , X_g ] = 0$ with (in an obvious notation) $g^i (x) =
B^i_j x^j + G^i$, then $[X_f , X_B ] = 0$ as well, i.e. $X_F$ is
in the centralizers of both $X_0$ and $X_B$. More generally, if
$X_f$ is an element of some Lie algebra ${\cal G}$, then under suitable conditions, see \cite{CiGa}, it can be put in joint normal form, i.e. (again with obvious notations) $[X_f , X_{B_i} ] = 0$.  $\odot$

\subsection{Sporadic resonances and invariance relations}

Let us consider again the resonance equation (2). It is clear that
if there are non-negative integers $\s_i$ (some of them nonzero) such that 
$$ \sum_{i=1}^n \, \s_i \la_i \ = \ 0 \ , \eqno(5) $$
then we always have infinitely many resonances. The monomial $\phi
= x^\s$ will be called a {\it resonant scalar monomial}. It is an invariant of $X_0$, and any multiindex $\mu$
with $\mu_i = k \s_i + \de_{i \a}$ provides a resonance relation
$(\mu \cdot \la ) = \la_\a$ related to the eigenvalue $\la_\a$; in
other words, any monomial $x^{k \s} x^\a = \phi x^\a$ is resonant,
and so is any vector $\vb_{k \s + e_\a , \a}$.

Therefore, we say that (5) identifies a {\it invariance relation}.
The presence of invariance relations is the only way to have
infinitely many resonances in a finite dimensional system (see \cite{enf}).

Any nontrivial resonance (2) such that there is no $\s$ with $\s_i \le \mu_i$ (for all $i=1,...,n$) providing an invariance relation, is said to be a {\it sporadic resonance}. Sporadic resonances are always in finite number (if any) in a finite dimensional system \cite{enf}.

Any invariance relation (5) such that there is no $\nu$ with
$\nu_i \le \s_i$ (and of course $\nu \not= \s$) providing another
invariance relation, is said to be an {\it elementary invariance
relation}. Every invariance relation is a linear combination (with nonnegative integer coefficients) of elementary ones. Elementary invariance relations are always in finite number (if any) in a finite dimensional system \cite{enf}.

\section{Embedding systems with invariance relations in quasi-linear systems}

In this section we will discuss how the procedure described in
\cite{con} generalizes, in connection with a ``splitting'' of the
system described by $Y$ \cite{split}, in the presence of
invariance relations.

We should preliminarily identify all sporadic resonances $(\mu
\cdot \la ) = \la_\a$ and elementary invariance relations $(\s
\cdot \la ) = 0$. We associate resonant monomials $x^\mu$ and
resonant vectors $\vb_{\mu,\a}$ to the former ones, and invariant
monomials $x^\s$ to the latter ones.

We then introduce two set of new coordinates: these will be the
coordinates $w^1 , ... , w^r$ in correspondence with sporadic
resonances (as in \cite{con}), and other new coordinates $\phi^1,
... , \phi^m$ in correspondence with elementary invariance
relations.

We should also assign evolution equations for the $w$ and $\phi$ coordinates; these will be given in agreement with (1) itself. That is, the equations for the $w$ will be
$$ {\d w^j \over \d t} \ = \ {\pa w^j \over \pa x^i} \, {\d x^i \over \d t} \ := \ h^j (x,w,\phi) \ ; \eqno(6) $$
and as for the $\phi$'s we assign
$$ {\d \phi^a \over \d t} \ = \ {\pa \phi^a \over \pa x^i} \, {\d x^i \over \d t} \ := \ z^a (x,w,\phi) \ . \eqno(7) $$

We will thus consider the enlarged space $W = (x,w,\phi) =
\R^{n+r+m}$, and in this the vector field
$$ Y \ = \ f^i (x,w,\phi) \, {\pa \over \pa x^i} \ + \ h^j (x,w,\phi) \, {\pa \over \pa w^j} \ + \ z^a (x,w,\phi) \, {\pa \over \pa \phi^a} \ . \eqno(8) $$

Note that some ambiguity is present here, in that we can write the
coefficients of this vector field in different ways as a function
of the $x,w,\phi$. Indeed, the vector field $Y$ is uniquely
defined only on the manifold identified by $\psi^i := w^i -
x^{\mu_{(i)}} = 0$, $\phi_a - \zeta_a (x) = 0$.

\medskip\noindent
{\bf Lemma 1.} {\it The $(n+m)$-dimensional manifold $M \ss W$ identified by $\psi^i := w^i - x^{\mu^{(i)}} = 0$ is invariant under the flow of $Y$.}

\medskip\noindent
{\bf Proof.} Obvious by construction. $\triangle$

\medskip\noindent
{\bf Lemma 2.} {\it The functions $z^a$ defined in (7) can be written in terms of the $\phi$ variables alone, i.e. $\pa z^a / \pa x^i = \pa z^a / \pa w^j = 0$.}

\medskip\noindent
{\bf Proof.} Every analytic invariant of $X_0$ can be represented as a convergent series in the $z^a$ (see \cite{enf}); their evolution is also invariant under the $X_0$ action, hence can also be written in terms of invariants, hence of the $z^a$ themselves. $\triangle$

\medskip\noindent
{\bf Corollary 1.} {\it The evolution of the $\phi$ variables is described by a (nonlinear) equation in the $\phi$ variables only.}

\medskip\noindent
{\bf Proof.} This is merely a restatement of lemma 2 above. Note that the equations for $x$ and $w$ depend on $\phi$ and are therefore nonautonomous. $\triangle$

\medskip\noindent
{\bf Proposition 3.} {\it The analytic functions $f^i$ and $h^j$
defined above can be written as linear in the $x$ and $w$
variables, the coefficients being functions of the $\phi$ variables.}

\medskip\noindent
{\bf Proof.} Recall each $w^j$ is a monomial $w^j =
x_1^{p_1} ... x_n^{p_n}$ with $(p , \la) = \la_s$ for some
$s=1,...,n$; with reference to this integer $s$, we add a label to
$w_j$, i.e. write $w_j^{(s)}$. 
By construction and by the results above, each $f^m$ can be written in 
the form $f^m = a^m (\phi) \cdot x_m + \sum_k c_k^m (\phi) w_k^{(m)}$, with analytic $a_m$ and $c_k^m$. So the assertion for the $f^m$ is obvious.

The time derivative of $w_j^{(s)}$ under the flow of (1) will be ${\dot w}_j^{(s)} = (\pa w_j^{(s)} / \pa x^m ) f^m$. 
Therefore it is sufficient to show that $(\pa w_j^{(s)}/\pa x^m ) w_k^{(m)}$ is zero or a multiple of $x^m$ or of some $w_\ell^{(s)}$ with a suitable resonant scalar monomial as a factor. But the above operation just means to replace one factor $x^m$ by $w_k^{(m)}$, and the resulting linear combination of the eigenvalues still yields a resonance relation (2) with $\la_s$ on the right hand side. See \cite{seb} for a different approach to the proof. $\triangle$

\medskip\noindent
{\bf Corollary 2.} {\it The evolution of the $x$ and $w$ variables is described by nonautonomous linear equations, obtained by inserting the solution $\phi = \phi (t)$ of the equations for $\phi$ in the general equations $\xd = f (x,w,\phi)$, ${\dot w} = h (x,w,\phi)$.}
\medskip

\medskip\noindent
{\bf Proof.} Obvious. $\triangle$
\medskip

\medskip\noindent
{\bf Remark 3.} We will also say that the vector field $Y$ is quasi-linear, meaning by this that it is linear in the $x$ and $w$ variables. In this way we recover -- as a special case -- the situation discussed in \cite{Lie} as well as the terminology used there. $\odot$

\medskip\noindent
{\bf Remark 4.} The results obtained here extend and unify those given in \cite{split,con}; see also \cite{enf}. As the $\phi$ identify group orbits for the group $G$ generated by the Lie algebra, we interpret $\dot\phi = z (\phi)$ as an equation in orbit space, and the equation for $(x,w)$ as an equation on the Lie group $G$. Methods for the solution of the latter are discussed in \cite{WN}, see also \cite{CGM}. $\odot$

\medskip\noindent
{\bf Remark 5.} If no invariance relations are present, hence no
$\phi$ variables are introduced, then the system describing the
time evolution of the $x,w$ variables is linear; this is the
situation studied in \cite{con}. Note that in this case we have
exactly the interpretation of normal forms as projection of a
linear system to an invariant manifold, without symmetry
reduction. $\odot$

\medskip\noindent
{\bf Remark 6.} If there are no sporadic resonances of order greater than one then Proposition 3 yields a linear system for the $f^i$, with functions of the $\phi$ variables as coefficients. Therefore, upon solving the reduced equation for the $\phi$ variables one obtains a non-autonomous linear system. Moreover, if all eigenvalues are distinct then we have a product system of one-dimensional equations. $\odot$

\medskip\noindent
{\bf Remark 7.} Finally, we note that if $\phi (t)$ converges to some $\phi_0$ (this is always the case if the $\phi$ space is one-dimensional and $|\phi (t)|$ does not escape to infinity), the asymptotic evolution of the system is governed by a linear autonomous equation for $x$ and $w$ (see \cite{Thi} for the behavior of 
asymptotically autonomous equations). Similarly, if there is a periodic solution $\bar\phi (t)$ with $\phi (t) \to \bar\phi (t)$, the asymptotic evolution of the system is governed by a linear equation with periodic coefficients for $x$ and $w$. $\odot$

\section{Examples}

{\bf Example 1.} (See \cite{con}). For $A = {\rm diag} (1,k)$, $k \in {\bf N}$, the only resonant vector is $\vb = x^k \eb_2$, corresponding to a sporadic resonance, and there is no invariance relation. Systems in normal form correspond to
$$ X = x \pa_x + (k y + c x^k) \pa_y $$
with $c$ a real constant. According to our procedure, we define $w = x^k$, and obtain
$$ Y \ = \ x \pa_x + (k y + c w) \pa_y + k w \pa_w \ ; $$
the invariant manifold $M$ is given by $\psi := w - x^k = 0$. The
solution to the system in $W$ for initial data $(x_0,y_0,w_0)$ is
$ x(t) = x_0 e^t$, $y(t) = y_0 e^{kt} + (c_1 k w_0) t e^{kt}$,
$w(t) = w_0 e^{kt}$; for initial data on $M$, i.e. $w_0 = x_0^k$,
the solution remains on $M$ and its projection to $\R^2 = (x,y)$
is $ x(t)= x_0 e^t$, $y (t) = [ y_0 + (c_1 k x_0^k) t ] e^{kt} $.

\medskip\noindent
{\bf Example 2.} Consider the matrix
$$ A \ = \ \pmatrix{ 0 & -1 & 0 \cr 1 & 0 & 0 \cr 0 & 0 & 1 \cr} $$
with eigenvalues $(-i,i,1)$. There is one elementary invariance
relation, $\la_1 + \la_2 = 0$, and no sporadic resonance. The linear centralizer is spanned by $A$ itself and by matrices $ D_1 = {\rm diag} (1,1,0)$ and $D_2 = {\rm diag} (0,0,1)$. The ring of
invariants is generated by $r^2 := x^2 + y^2$. Systems in normal
forms are written as
$$ \begin{array}{l}
\xd = \a (r^2) x - \b (r^2) y \\
\yd = \b (r^2) x + \a (r^2) y \\
{\dot z} = \ga (r^2) z \end{array} $$
where $\a , \b , \ga $ are arbitrary power series.

Following our procedure, we introduce one further variable $\phi = r^2$; for this we have $d \phi / d t = 2 (x \xd + y \yd ) = 2 r^2 \a (r^2 ) $. Hence the system in $W = \R^4$ is written as
$$ \begin{array}{l}
\xd \ = \ \a (\phi) \, x \ - \ \b (\phi) \, y \\
\yd \ = \ \b (\phi) \, x \ + \ \a (\phi) \, y \\
{\dot z} \ = \ \ga (\phi) \, z \\
{\dot \phi} = \ 2 \, \phi \, \a (\phi) \ . \end{array} $$

\medskip\noindent
{\bf Example 3.} Consider the matrix
$$ A \ = \ \pmatrix{ 0 & -1 & 0 & 0 \cr 1 & 0 & 0 & 0 \cr  0 & 0 & 1 & 0 \cr 0 & 0 & 0 & k \cr} $$
($k \in {\bf N}$, $k > 1$) with eigenvalues $(-i,+i,1,k)$. There
is one sporadic resonance, $k \la_3 = \la_4$, and one elementary
invariance relation, $\la_1 + \la_2 = 0$. The linear centralizer
is spanned the  matrices $D_1 = {\rm diag} (1,1,0,0)$, $D_2 = {\rm
diag} (0,0,1,0)$, $D_3 = {\rm diag} (0,0,0,1)$ together with
$$ M_1 = \pmatrix{0&-1&0&0\cr 1&0&0&0\cr 0&0&0&0\cr 0&0&0&0\cr} \ \ {\rm and} \ \ M_2 = \pmatrix{0&0&0&0\cr 0&0&0&0\cr 0&0&0&0\cr 0&0&k&0\cr} \ . $$
Systems in normal forms are written as
$$ \begin{array}{l}
\xd_1 = \a (r^2) x_1 - \b (r^2) x_2 \\
\xd_2 = \b (r^2) x_1 + \a (r^2) x_2 \\
\xd_3 = \ga (r^2) x_3 \\
\xd_4 = \eta (r^2) x_4 + \theta (r^2) x_3^k \end{array} $$
where $r^2 := x^2 + y^2$ and $\a , \b , \ga , \eta , \theta $ are arbitrary power series.

Following our procedure we introduce $\phi = r^2$, for which $d \phi / d t = 2 (x \xd + y \yd ) = 2 r^2 \a (r^2 ) $, and $w = x_3^k$ for which $d w / d t = k x_3^{k-1} \xd_3 = k \ga (r^2) x_3^k$. Hence the system in $W = \R^5$ is written as
$$ \begin{array}{l}
\xd_1 = \a (\phi) x_1 - \b (\phi) x_2 \\
\xd_2 = \b (\phi) x_1 + \a (\phi) x_2 \\
\xd_3 = \ga (\phi) x_3 \\
{\dot w} = k \ga (\phi) w \\
{\dot \phi} = 2 \phi \a (\phi) \ . \end{array} $$

\section{Perturbed oscillators}

{\bf Example 4. Perturbation of oscillators in 1:1 resonance.}
Consider the matrix
$$ A \ = \ \pmatrix{ 0 & -1 & 0 & 0 \cr 1 & 0 & 0 & 0 \cr  0 & 0 & 0 & -1 \cr 0 & 0 & 1 & 0 \cr} \ , $$
which we also write in block form as
$$ A \ = \ \pmatrix{ J & 0 \cr 0 & J \cr} \ , \ {\rm where} \ J \ = \ \pmatrix{0&-1\cr 1&0\cr} \ , $$
with eigenvalues $(-i,+i,-i,i)$. There are no sporadic resonances of order greater than one, and four elementary invariance relations:
$$ \la_1 + \la_2 = 0 \ , \ \la_3 + \la_4 = 0 \ , \
\la_1 + \la_4 = 0 \ , \ \la_2 + \la_3 = 0 \ ; $$
all other resonances can be described in terms of these. 
We stress that the equations describing these invariance relations are
linearly dependent; however they should be considered, according
to our definition, as different elementary ones. Corresponding to
this, the associated invariant quantities will not be functionally
independent (obviously, we cannot have more than three independent
invariants for a flow in $\R^4$).

The linear centralizer of $A$ is an eight-dimensional algebra, spanned by the following matrices (in two by two block notation):
$$\begin{array}{l}
B_1 = \pmatrix{I&0\cr 0&0\cr} \ , \
B_2 = \pmatrix{0&0\cr 0&I\cr} \ , \
B_3 = \pmatrix{0&I\cr I&0\cr} \ , \
B_4 = \pmatrix{0&J\cr -J&0\cr} \ , \\
~ \\
S_1 = \pmatrix{J&0\cr 0&0\cr} \ , \
S_2 = \pmatrix{0&0\cr 0&J\cr} \ , \
S_3 = \pmatrix{0&I\cr -I&0\cr} \ , \
S_4 = \pmatrix{0&J\cr J&0\cr} \ .
\end{array} \eqno(9) $$
Note here we have chosen a basis with $B_i = B_i^+$, $S_i = - S_i^+$.

The linear system $\.\xi = A \xi$ describes two oscillators in 1:1 resonance; the normal form will correspond to a perturbation of these, generically breaking the exchange symmetry among the two oscillators.
Systems in normal form are compactly written as
$$
\pmatrix{ {\dot x}\cr {\dot y}\cr {\dot z}\cr {\dot w}\cr} \ = \
\pmatrix{
\a & - \b & \ga & - \eta \cr
\b & \a & \eta & \ga \cr
\mu & - \nu & \s & - \tau \cr
\nu & \mu & \tau & \s \cr}
\ \pmatrix{ x\cr y\cr z\cr w\cr} $$
where $\a , \b , ... , \tau$ are arbitrary power series in the elementary invariants
$$ \phi_1 = x^2 + y^2 \ , \ \phi_2 = z^2 + w^2 \ , \ \phi_3 = xz + yw \ , \ \phi_4 = xw - yz \ ; $$
note that $\phi_a = (\xi , B_a \xi)$, with $(.,.)$ the scalar product.
We abbreviate the above evolution equation as
$$ {\dot \xi} \ = \ K (\phi) \ \xi \ . $$
The evolution equations for the $\phi$, as required by our procedure, are simply (only the selfadjoint matrices defined in (9) appear)
$$ {\dot \phi}_a = \( \xi , (B_a K + K^+ B_a ) \xi \) \ . $$

\medskip\noindent
{\bf Example 5. Perturbation of oscillators in $1:k$ resonance.}
Consider the matrix
$$ A \ = \ \pmatrix{ 0 & -1 & 0 & 0 \cr 1 & 0 & 0 & 0 \cr  0 & 0 & 0 & -k \cr 0 & 0 & k & 0 \cr} $$
(with $k \in {\bf N}$, $k > 1$) with eigenvalues $(-i,+i,-i
k,ik)$. This is put in diagonal form passing to
variables
$$ \xi_1 = (x_1 - i x_2)/2 \ , \ \xi_2 = (x_1 + i x_2)/2 \ , \ 
\xi_3 = (x_3 - i x_4)/2 \ , \ \xi_4 = (x_3 + i x_4)/2 \ , $$
which we use in intermediate computations below.
We also write $ \xi = \Lambda x$, $x = \Lambda^{-1} \xi$, with
$$ \Lambda = {1 \over 2} \ \pmatrix{1&-i&0&0\cr 1&1&0&0\cr 0&0&1&-i\cr 0&0&1&i\cr} \ ; \ \Lambda^{-1} = \pmatrix{1&1&0&0\cr i&-i&0&0\cr 0&0&1&1\cr 0&0&i&-i\cr} \ . $$

There are four sporadic resonances:
$$ k \la_1 = \la_3 \ ,  \ k \la_2 = \la_4 \ , \
\la_3 + (k-1) \la_2 = \la_1 \ ,  \ \la_4 + (k-1) \la_1 = \la_2 \ . $$
Moreover, there are four elementary invariance relations:
$$ \la_1 + \la_2 = 0 \ , \ \la_3 + \la_4 = 0 \ , \
k \la_1 + \la_4 = 0 \ , \ k \la_2 + \la_3 = 0 \ . $$
All other resonances can be described in terms of these. 

Systems in normal forms are written as
$$ \begin{array}{l}
\.\xi_1 = \a_1 \xi_1 + \vth_1 \xi_2^{k-1} \xi_3 \\
\.\xi_2 = \a_2 \xi_2 + \vth_2 \xi_1^{k-1} \xi_4 \\
\.\xi_3 = \a_3 \xi_3 + \vth_3 \xi_1^k \\
\.\xi_4 = \a_4 \xi_4 + \vth_4 \xi_2^k \ ,
\end{array} \eqno(10) $$
where $\a_i , \vth_i $ are arbitrary power series in the invariants of the linear flow.

Following our procedure, we introduce variables
$$ w_1 = \xi_1^k \ , \ w_2 = \xi_2^k \ , \
w_3 = \xi_2^{k-1} \xi_3 \ , \ w_4 = \xi_1^{k-1} \xi_4 \ ,  $$
related to sporadic resonances. We also introduce variables
related to elementary invariance relations, given by
$$ \phi_1 = \xi_1 \xi_2 \ , \ \phi_2 = \xi_3 \xi_4 \ , \
\phi_3 = \xi_1^k \xi_4 \ , \ \phi_4 = \xi_2^k \xi_3 \ . $$

We must then introduce evolution equations for the $w$ and $\phi$ variables according to our procedure, i.e. according to (6) and (7) above.
As for the $w$, we get
$$ \begin{array}{l}
\.w_1 = k \a_1 w_1 + \vth_1 \phi_1^{k-1} \xi_3 \\
\.w_2 = k \a_2 w_2 + \vth_2 \phi_1^{k-1} \xi_4 \\
\.w_3 = [\a_3 + (k-1) \a_2 ] w_3 + [ (k-1) \vth_2 \phi_1^{k-2} \phi_2 + \vth_3 \phi_1^{k-1} ] \xi_1 \\
\.w_4 = [\a_4 + (k-1) \a_2 ] w_4 + [ (k-1) \vth_1 \phi_1^{k-2} \phi_2 + \vth_4 \phi_1^{k-1} ] \xi_2 \ .
\end{array}$$

Let us now consider the equations for the $\phi$; we easily get
$$ \begin{array}{l}
\.\phi_1 = (\a_1 + \a_2 ) \phi_1 + \vth_2 \phi_3 + \vth_1 \phi_4 \\
\.\phi_2 = (\a_3 + \a_4 ) \phi_2 + \vth_3 \phi_3 + \vth_4 \phi_4 \\
\.\phi_3 = (k \a_1 + \a_4 ) \phi_3 + k \vth_1 \phi_1^{k-1} \phi_2 + \vth_4 \phi_1^k \\
\.\phi_4 = (k \a_2 + \a_3 ) \phi_4 + k \vth_2 \phi_1^{k-1} \phi_2 + \vth_3 \phi_1^k \ .
\end{array} \eqno(11) $$

Summarizing, all systems of the form (10) -- i.e. in normal form with respect to the linear part $\.\xi = A \xi$ -- are written as the autonomous system (11) for the $\phi$ variables, plus a linear nonautonomous system, which introducing the notation $\eta = (\xi ; w)$ can be written as
$$ \.\eta \ = \ M \ \eta $$
where $M = M (\phi)$ is a matrix which we write as $M = D + L$, where
$$ D \ = \ {\rm diag} (\a_1 , \a_2 , \a_3 , \a_4 ; k \a_1 , k \a_2 , \a_3 + (k-1) \a_2 , \a_4 + (k-1) \a_1 ) $$
and $L$ is an off-diagonal sparse matrix with nonzero terms
$$ \begin{array}{c}
L_{17} = \vth_1 \ , \ L_{28} = \vth_2 \ , \ L_{35} = \vth_3 \ , \ L_{46} = \vth_4 \ ; \ 
L_{53} = \vth_1 \phi_1^{k-1} \ , \ L_{64} = \vth_2 \phi_1^{k-1} \ , \\
L_{71} = [ (k-1) \vth_2 \phi_1^{k-2} \phi_2 + \vth_3 \phi_1^{k-1} ] \ , \ 
L_{82} = [ (k-1) \vth_1 \phi_1^{k-2} \phi_2 + \vth_4 \phi_1^{k-1} ] \ .
\end{array}
$$

\medskip\noindent
{\bf Example 6. Perturbation of two oscillators with no resonance.} Consider the matrix 
$$ A \ = \ \pmatrix{ 0 & -1 & 0 & 0 \cr 1 & 0 & 0 & 0 \cr  0 & 0 & 0 & -\pi \cr 0 & 0 & \pi & 0 \cr} $$
with eigenvalues $(-i,+i,-\pi i ,\pi i)$.

Now there are no sporadic resonances of order greater than one, and two elementary
invariance relations:
$$ \la_1 + \la_2 = 0 \ , \ \la_3 + \la_4 = 0 \ . $$
Note that now we have only two invariants, $r_1^2 = x_1^2 + x_2^2$ and $r_2^2 = x_3^2 + x_4^2$: this is easily understood as we have irrational flow on the two-torus ${\bf T}^2 \ss \R^4$.
The centralizer of $A$ corresponds to matrices
$$ M \ = \ \pmatrix{ a & - b & 0 & 0 \cr b & a & 0 & 0 \cr 0&0& c & - d \cr 0&0& d & c \cr} $$
Thus systems in normal form will be written as
$$ \pmatrix{\xd_1 \cr \xd_2 \cr \xd_3 \cr \xd_4 \cr} \ = \ \pmatrix{ \a & - \b & 0 & 0 \cr \b & \a & 0 & 0 \cr 0&0& \ga & - \eta \cr 0&0& \eta & \ga \cr} \ \pmatrix{x_1 \cr x_2 \cr x_3 \cr x_4 \cr} $$
with $\a , \b , \ga , \eta$ being power series in $r_1^2 , r_2^2$.

Following our procedure we introduce variables $ \phi_1 = x_1^2 + x_2^2$ and $\phi_2 = x_3^2 + x_4^2 $, whose evolution is given by
$$ {\dot \phi}_1 \ = \ 2 \, \a \, \phi_1 \ , \ {\dot \phi}_2 \ = \ 2 \, \ga \, \phi_2 \ . $$

\medskip\noindent
{\bf Example 7. Perturbation of oscillators in 1:1:1 resonance.}
Consider the six-dimensional matrix written in block form as
$$ A \ = \ \pmatrix{\om J & 0 & 0 \cr 0 & \om J & 0 \cr 0 & 0 & \om J \cr}  \ \ {\rm where} \ \ J \ = \ \pmatrix{0 & - 1 \cr 1 & 0 \cr} \ . $$
Passing to coordinates $(\xi_j,\eta_j) = (p_j+i q_j, p_j- i q_j)$,
this reads
$$ \^A \ = \ {\rm diag} (i \om , - i \om , i \om , - i \om , i \om , - i \om ) \ . $$
The eigenvalues are $\la_k \ = \ (-1)^k i \om $, hence there is no sporadic resonance of order greater than one\footnote{Note these are present in the case of $1:k:\ell$ resonance (with integers $k,\ell > 1$).} and nine invariance relations:
$$ \begin{array}{lll}
\la_1 + \la_2 = 0 & \la_3 + \la_4 = 0 & \la_5 + \la_6 = 0 \\
\la_1 + \la_4 = 0 & \la_1 + \la_6 = 0 & \la_3 + \la_6 = 0 \\
\la_2 + \la_3 = 0 & \la_2 + \la_5 = 0 & \la_4 + \la_5 = 0
\end{array} $$ The invariants corresponding to these are of course
$$ \begin{array}{lll}
\^\phi_1 = \xi_1 \eta_1 , & \^\phi_2 = \xi_2 \eta_2 , & \^\phi_3 = \xi_3 \eta_3 , \\
\^\phi_4 = \xi_1 \eta_2 , & \^\phi_5 = \xi_1 \eta_3 , & \^\phi_6 = \xi_2 \eta_3 , \\
\^\phi_7 = \xi_2 \eta_1 , & \^\phi_8 = \xi_3 \eta_2 , & \^\phi_9 =
\xi_3 \eta_2 .
\end{array} $$
Going back to the original coordinates, expressions for these are
recovered from (no sum on repeated indices)
$$ \xi_j \eta_j \ = \ (p_j^2 + q_j^2) \ \ ; \ \ \xi_j \eta_k \ = \ (p_j p_k + q_j q_k) + i (q_j p_k - p_j q_k) $$
and is thus more convenient to pass to a different basis for
invariant functions, i.e.
$$ \begin{array}{l}
\phi_1 := \^\phi_1 = (p_1^2 + q_1^2) \ \ , \ \ \phi_2 := \^\phi_2
= (p_2^2 + q_2^2) \ \ , \ \
\phi_3 := \^\phi_3 =  (p_3^2 + q_3^2) \ ; \\
\phi_4 := (\^\phi_4 + \^\phi_7)/2 = (p_1 p_2 + q_1 q_2) \ \ , \ \
\phi_5 := (\^\phi_4 - \^\phi_7)/(2i) = (q_1 p_2 - p_1 q_2) \ ; \\
\phi_6 := (\^\phi_5 + \^\phi_9)/2 = (p_1 p_3 + q_1 q_3) \ \ , \ \
\phi_7 := (\^\phi_5 - \^\phi_9)/(2i) = (q_1 p_3 - p_1 q_3) \ ; \\
\phi_8 := (\^\phi_6 + \^\phi_8)/2 = (p_2 p_3 + q_2 q_3) \ \ , \ \
\phi_9 := (\^\phi_6 - \^\phi_8)/(2i) = (q_2 p_3 - p_2 q_3) \ .
\end{array} $$

The centralizer $C (A)$ of $A$ in ${\rm Mat} (6,R)$ is an algebra
spanned by eighteen matrices; these are written in $2 \times 2$
block form as
$$ \pmatrix{C_{11} & C_{12} & C_{13} \cr C_{21} & C_{22} & C_{23} \cr
C_{31} & C_{32} & C_{33} \cr} $$ where the $C_{ij}$ are 
two-dimensional matrices written as ($\a_{ij}$ and $\b_{ij}$ real
constants)
$$ C_{ij} \ = \ \a_{ij} \, I \ + \ \b_{ij} \, J \ . $$
It is easy to extract from these a basis made of nine selfadjoint
matrices $B_k = B_k^+$ and nine antiselfadjoint ones $S_k = -
S_k^+$. With $\xb = (p_1,q_1,p_2,q_2,p_3,q_3)$, and $(.,.)$ the
standard scalar product in $\R^6$ these can be chosen so that $
\phi_a = (\xb , B_a \xb) $.

In this compact notation the $1:1:1$ resonant three-dimensional
oscillator is described by ${\dot \xb} = A \xb$. We write the
normal form of evolution equations for perturbation of this as
$$ \begin{array}{ll}
{\dot \xb} \ &= \ K (\phi ) \ \xb \\
{\dot \phi}_a \ &= \( \xb \, , \, (B_a K + K^+ B_a ) \xb \right)
\end{array} $$ with $K$ an arbitrary matrix in $C(A)$.

\section{Bifurcations}

{\bf Example 8. Hopf bifurcation.} Consider the matrix
$$ A \ = \ \pmatrix{0&-\om_0\cr \om_0&0\cr} $$
with eigenvalues $\la_1 = - i \om_0$, $\la_2 = i \om_0$. There is
no sporadic resonance, and one elementary invariance relation, $\la_1 + \la_2 = 0$, with associated invariant $\phi = x^2 + y^2$.
The most general system in normal form is therefore
$ \xd = \a (\phi ) x - \b (\phi ) y$, $\yd = \b (\phi ) x + \a (\phi ) y $. According to our procedure, the evolution equation for the new coordinate $\phi$ will be $ \.\phi = 2 \phi \a (\phi)$. 
As the linear part of the system is given by $A$, we must require $\a (0) = 0 $, $\b (0) = \om_0$.

In applications, one is interested in the case where the system does also depend on an external (``control'') parameter $\mu$, which usually does not evolve in time\footnote{A different framework is provided by dynamic bifurcations \cite{Ben,Nei}.}, the linear part being given by $A$ at the critical value. In this case the normal form and the $\phi$ evolution equation read
$$\begin{array}{l}
\xd \ = \ \a (\phi , \mu ) \, x \ - \ \b (\phi , \mu) \, y \\
\yd \ = \ \b (\phi , \mu ) \, x \ + \ \a (\phi , \mu) \, y \\
\.\phi \ = \ 2 \, \phi \, \a (\phi , \mu) \ .
\end{array} $$

In the standard model of Hopf bifurcation, $\a (\phi , \mu ) = \mu - c \phi $, and we write $\b (\phi , \mu ) = \om_0 + b (\phi , \mu)$ with $b(0,0) = 0$. This corresponds to the normal form
$$ \cases{
\xd = \mu x - \om_0 y - b (x^2+y^2 , \mu) y - c (x^2 + y^2 ) x & \cr
\yd = \om_0 x + \mu y + b (x^2+y^2 , \mu) x - c (x^2 + y^2 ) x & \cr} $$
which in our approach reads
$$ \cases{
\xd = \mu x - \om_0 y - b (\phi , \mu) y - c (\phi ) x & \cr
\yd = \om_0 x + \mu y + b (\phi , \mu) x - c (\phi ) x & \cr 
\.\phi  =  2 \, \phi \, \a (\phi , \mu) \ . & \cr}
$$
Note that the space of invariants is one-dimensional (with the additional constraint $\phi \ge 0$); thus, either $\phi (t)$ is unbounded for $t > 0$, or it approaches one of the zeroes of the function $\a (\phi , \mu)$, say $\phi_0$. In this case, the system reduces asymptotically to a linear one: 
$$ 
\pmatrix{\xd \cr \yd \cr} \ = \ \pmatrix{
0 & - \om_0 - b (\phi_0 , \mu ) \cr 
\om_0 + b (\phi_0 , \mu) & 0 \cr} \ \pmatrix{ x \cr y \cr} \ . $$
The standard analysis of Hopf bifurcation is readily recovered in this way. 
\bigskip

Note that we can also look at Hopf bifurcation in a slightly
different way, i.e. include the $\mu$ variable from the beginning.
In this case the matrix $A$ is given by
$$ A = \pmatrix{0&-1&0\cr 1&0&0\cr 0&0&0\cr} $$
with eigenvalues $(-i,i,0)$ and invariance relations $\la_1 +
\la_2 = 0 $ and $\la_3 = 0$; the associated invariants are $\phi$
and $\mu$.
The linear centralizer of $A$ is spanned by matrices
$$ M = \pmatrix{a&-b&0\cr b&a&0\cr 0&0&c\cr} $$
and correspondingly the normal form will be
$$ \begin{array}{l}
\xd = \a (\phi , \mu) x - \b (\phi , \mu) y \\
\yd = \b (\phi , \mu) x + \a (\phi , \mu) y \\
\.\mu = \ga (\phi , \mu)
\end{array} $$

The equation for $\phi$ is just the one given above, and we are
led to the same system; interpreting $\mu$ as an external control
parameter forces $\ga (\phi , \mu ) \equiv 0$ (if not, there is a feedback of the system on the control parameter).

\medskip\noindent
{\bf Example 9. Hamiltonian Hopf bifurcation.} Consider the matrix
$$ A \ = \ \pmatrix{\mu & - \om & 0 & 0 \cr \om & \mu & 0 & 0 \cr 0&0&- \mu &-\om \cr 0&0&\om & -\mu \cr} \ = \ \pmatrix{\mu I + \om J & 0 \cr 0 & - \mu I + \om J \cr} $$
with eigenvalues $\la_1 = -\mu - i \om$, $\la_2 = -\mu + i \om$, $\la_3 = \mu -i \om$, $\la_4 = \mu + i \om$. We assume $\om \not= 0$ (so it could be rescaled to $\om = 1$) and $\mu \not=0$; the case $\mu = 0$ corresponds to example 4. In applications, one considers the case where $\mu$ is an external control parameter and studies the changes as this is varied; $\mu = 0$ is a critical value.

The matrix $A$ is diagonalized, for all $\mu$, passing to variables $\xi^i$ as in example 5 above. The evolution $\xd = A x$ preserves the symplectic structure $\kappa = \d x^1 \wedge \d x^3 + \d x^2 \wedge \d x^4$.

It is easy to check, for generic $\mu$, that there is no sporadic resonance and that there are two elementary invariance relations, given by $\la_1 + \la_4 = 0$ and $\la_2 + \la_3 = 0$. The corresponding invariants will be $ \varphi_1 = \xi^1 \xi^4$, $\varphi_2 = \xi^2 \xi^3$; they are complex conjugate, and correspond to real invariants 
$ \phi_1 = x_1 x_3 + x_2 x_4$ and $\phi_2 = x_1 x_4 - x_2 x_3$. 
These are also written for later reference as $\phi_a = (1/2) (\xi , B_a \xi)$, with $B_a$ obvious four-dimensional symmetric matrices. 

The linear centralizer of $A$ (for $\mu \not= 0$) is given by matrices written in block form, with $\a_k$ and $\b_k$ real constants, as
$$ M \ = \ \pmatrix{
\a_1 I + \b_1 J & 0 \cr
0 & \a_2 I + \b_2 J \cr} \ . $$

Correspondingly, systems in normal form will be given by
$$ \xd \ = \ M \, x $$
where now $\a_k , \b_k $ will be functions of the
invariants $\phi_1 , \phi_2$. Note that such systems in
general do not preserve the symplectic form $\kappa$, unless $\a_2 = - \a_1 $ and $\b_2 = \b_1$. 
The evolution of the $\phi$ is given by
$ {\dot \phi}_a = (1/2) \( x , (M^+ B_a + B_a M ) x \)$. 

It is convenient to write the system in terms of the two-dimensional vectors $\eta_1 = (x^1,x^2)$, $\eta_2 = (x^3,x^4)$ and $\phi = (\phi_1,\phi_2)$. Moreover, we single out the linear part in the functions $\a_k$ and $\b_k$, writing $\a_k (\phi) = (-1)^{k+1} \mu + a (\phi)$, $\b_k (\phi) = \om + b_k (\phi)$ with $a_k (0) = b_k (0) = 0$; the smooth functions are $a_k$, $b_k$ are arbitrary apart from this constraint, and can also depend on the parameters $\mu$ and $\om$.

The system is hence described in our approach by the following equations: 
$$ \begin{array}{ll}
{\dot \eta}_1 =& (\mu I + \om J) \, \eta_1 \ + \ [a_1 (\phi) \cdot I \, - \, b_1 (\phi) \cdot J ] \, \eta_1 \ , \\
{\dot \eta}_2 =& (- \mu I + \om J) \, \eta_2 \ + \ [a_2 (\phi) \cdot I \, - \, b_2 (\phi) \cdot J ] \, \eta_2 \ , \\ 
{\dot \phi} =& \[ \( a_1 (\phi) + a_2 (\phi) \) \cdot I \, + \, \( b_2 (\phi) - b_1 (\phi) \) \cdot J \] \ \phi \ . \end{array} $$ 

If $a_k , b_k$ are such that the system is hamiltonian, the $\phi$ are always strictly invariant, and we are reduced to a linear system on each level set of $\phi$; if the $a_k,b_k$ are such that the system is not hamiltonian but the (two-dimensional) equation for the $\phi$ satisfies the condition for the existence of a limit cycle, as is often the case in bifurcation problems, then remark 7 applies.

Let us look more closely to the perturbation of the case $\mu = 0$; as recalled above, the analysis of this case can be recovered from example 4. More precisely, we can rescale time so that $\om = 1$, and allow the arbitrary functions appearing in the analysis of example 4 to also depend on the parameter $\mu$.

Alternatively, we can proceed as at the end of the previous example, and include the $\mu$ variable from the beginning. With a $2 \oplus 2 \oplus 1$ block notation, we have now
$$ A \ = \ \pmatrix{\mu I + \om J & 0 & 0 \cr 0 & - \mu I + \om J & 0 \cr 0 & 0 & 0 \cr} $$
with linear centralizer 
$$ M \ = \ \pmatrix{
\a_1 I + \b_1 J & 0 & 0 \cr
0 & \a_2 I + \b_2 J & 0 \cr
0 & 0 & c \cr} \ . $$
In the normal form, going back to the original time parametrization,  we have 
$$ \begin{array}{ll}
{\dot \eta}_1 =& (\mu I + \om J) \, \eta_1 \ + \ [\a_1 (\phi,\mu) \cdot I \, - \, \b_1 (\phi,\mu) \cdot J ] \, \eta_1 \ , \\
{\dot \eta}_2 =& (- \mu I + \om J) \, \eta_2 \ + \ [\a_2 (\phi,\mu) \cdot I \, - \, \b_2 (\phi,\mu) \cdot J ] \, \eta_2 \ , \\ 
{\dot \phi} =& \[ \( \a_1 (\phi,\mu) + \a_2 (\phi,\mu) \) \cdot I \, + \, \( \b_2 (\phi,\mu) - \b_1 (\phi,\mu) \) \cdot J \] \ \phi \ , \\
{\dot \mu} =& \gamma (\phi,\mu) \ . \end{array} $$ 
Again, interpreting $\mu$ as an external control parameter requires $\ga (\phi , \mu ) \equiv 0$.

\section*{Acknowledgements}

This work was started by discussions during a visit by SW in the
Department of Mathematics of Universit\`a di Milano, and while GG
was a guest at the DFG-Graduierten\-kol\-leg ``Hierarchie und
Symmetrie in Mathema\-ti\-schen Mo\-del\-len'' at RWTH Aachen. We
would like to thank these Institutions for support to our work. GG
also acknowledges partial support by Fondazione CARIPLO and by
GNFM-INdAM.

\end{document}